\documentclass[prd,twocolumn,showpacs,nofootinbib,amsfonts,amssymb,amsmath]{revtex4-1}

\usepackage{graphicx}
\usepackage{simplewick}
\usepackage{epstopdf}
\usepackage{cases}
\usepackage{soul} 

\def\be{\begin{equation}}
\def\ee{\end{equation}}
\def\bea{\begin{eqnarray}}
\def\eea{\end{eqnarray}}

\newcommand{\la}{\lambda}

\newcommand{\bear}{\begin{eqnarray}}

\newcommand{\eear}{\end{eqnarray}}

\newlength{\tskip}
\newbox\pippobox

\def\be{\begin{equation}}
\def\ee{\end{equation}}
\def\bea{\begin{eqnarray}}
\def\eea{\end{eqnarray}}
\def\bx{{\bf x}}

\def\bk{{\bf k}}

\def\a{\alpha}

\def\9{\nabla}

\def\s{\sigma}

\def\g{\gamma}

\def\b{\beta}

\def\d{\delta}
\def\dd{{\rm d}}

\def\p{\phi}

\def\nn{\nonumber}
\def\half{\frac12}
\def\le{\left}
\def\ri{\right}
\def\6{\partial}

\def\f{\frac}

\def\la{\langle}
\def\ra{\rangle}

\def\tld{\tilde}
\def\0{(0)}
\def\half{\f{1}{2}}
\def\>{\rightarrow}

\begin{document}

\title{{\bf Future CMB ISW-Lensing bispectrum constraints on modified gravity in the Parameterized Post-Friedmann formalism}}

\author{Bin Hu$^{1}$}

\author{Michele Liguori$^{1,2}$}

\author{Nicola Bartolo$^{1,2}$}

\author{Sabino Matarrese$^{1,2}$}

\affiliation{$~^{1}$INFN, Sezione di Padova, Via Marzolo 8, 35131 Padova, Italy
\\$~^{2}$Dipartimento di Fisica e Astronomia ``G. Galilei", Universit\`a degli Studi di Padova , Via Marzolo 8, 35131 Padova, Italy}
\date{\today}

\begin{abstract}
We forecast the constraints on both Hu-Sawicki model and Bertschinger-Zukin model of $f(R)$ modified gravity within the Parameterized Post-Friedmann (PPF) formalism for the {\it Planck} satellite experiment by performing the joint analysis of ISW-Lensing bispectrum and CMB power spectrum. 
We find that, even considering the temperature-temperature mode of CMB power spectrum only, {\it Planck} data are expected to reduce the error bars on the modified gravity parameter $B_0$ (related to the present value of Compton wavelength of the extra scalar degree of freedom) at least one order magnitude compared with WMAP. The spectrum-bispectrum joint analysis can further improve the results by a factor ranging from $1.14$ to $5.32$ depending on the specific modified gravity model. One of our main results is that the cross-correlation between ISW-Lensing bispectrum and power spectrum can be safely neglected when performing the joint analysis. For simplicity, we only investigate the likelihood of one parameter ({$B_0$}) and fix all other cosmological parameters to their best-fit values in WMAP7yr results. 
\end{abstract}


\maketitle

\section{introduction}
Cosmic acceleration can arise from either an exotic form of energy with negative pressure, namely ``dark energy", or a modification of gravity that appears on large scales. 
As shown in \cite{Song:2006ej}, every expansion history that can be parameterized by
a dark energy model with $\rho_{{\rm DE}}(\ln a)$ can be reproduced by a one
parameter family of $f(R)$ models, which approaches
the Einstein-Hilbert action in the high curvature limit.
Hence, the observations of the cosmic background expansion cannot help us to distinguish 
the viable modified gravity from dark energy models. 
However, the dynamical evolution of fluctuations can provide us a tool to test gravity theory (e.g. \cite{Zhang:2007nk,Wang:2007fsa,Jain:2007yk,Uzan:2006mf,
Zhang:2005vt,Camera:2011ms,Martinelli:2011wi,Schmidt:2008hc,Calabrese:2009tt,
Calabrese:2008rt,Camera:2009uz,Wang:2010ug,GilMarin:2011xq}
and references therein). The main difference between modified gravity and dark energy models is the anisotropic stress on large scales. In most of the dark energy models the large scale anisotropic stress vanishes, while, in modified gravity models usually it does not. The deeper physical reason for this phenomenon is that in  modified gravity usually the evolution of the Newtonian potential $\Psi$ deviates from that of the spatial curvature potential 
$\Phi$. 

A viable modified gravity model should at least satisfy the following three conditions: first, it should pass the stringent tests in the local solar-system; second, it should respect the standard expansion history and accelerate the late-time expansion; third, it should avoid modifying the physics at recombination, which has been investigated by various CMB experiments with a high precision. 

The studies of modified gravity models, in principle, can be classified into two different frameworks (see, e.g., the discussion in Ref.~\cite{Bertschinger:2008zb}). The first is a model dependent method. One can start from a specific Lagrangian, investigating its dynamical behavior to finally give its predictions. Various viable modified gravity models have been proposed (see the review article \cite{Clifton:2011jh}). In this paper we mainly focus on $f(R)$ models (see~\cite{DeFelice:2010aj} for a review), such as Starobinsky model \cite{Starobinsky:1980te}, Hu-Sawicki (HS) model \cite{Hu:2007nk}. The other method is inspired by the Parameterized Post-Newtonian (PPN) approach for the solar-system test. In this case one can try to build a more or less model independent framework, in which many modified gravity models can be parameterized in an unified way.
The simplest idea is to directly generalize Eddington parameter ($\gamma\equiv\Phi/\Psi$) \cite{Edd} in static solar-system to a one unknown function of space and time $\gamma(t,\bx)$ in Friedmann universe.
Many studies (e.g., \cite{Bertschinger:2008zb,Amendola:2007rr,Zhao:2008bn,Zhao:2009fn,Hojjati:2011ix,Giannantonio:2009gi}) based on this framework show that this parameterization works quite well with the large scale structure data. 
However, as pointed in \cite{Bertschinger:2006aw}, the horizon-scale and superhorizon evolution need to be consistent with the background expansion of the universe. This is one of the reason why the simple $\gamma$ parameter is not sufficient when dealing with, such as ISW effect. Furthermore, as stated in \cite{Zuntz:2011aq} using a parameterization with insufficient
freedom significantly tightens the apparent theoretical constraints.
Still in the same spirit of the standard solar-system PPN approach, another viable parameterization has been proposed, the so called Parameterized Post-Friedmann (PPF) framework. The idea of PPF framework is to parameterize the background/linear fluctuation gravitational equations, which is firstly proposed in \cite{Tegmark:2001zc}\footnote{We are grateful to Tessa Baker for pointing this out}.After that different formalism in the domain of PPF have been proposed, such as Hu-Sawicki PPF formalism 
\cite{Hu:2007pj,Hu:2008zd,Fang:2008sn} that bridges between different scale regimes. In details, PPF allows to describe (and match) three different regimes of gravity modifications to GR, namely the super-horizon regime, the (sub-horizon) quasi-static regime (with an intermediate regime between the two) and a non-linear regime on smaller scales. Aside that Barker-Ferreira-Skordis-Zuntz also give their algorithm in \cite{Baker:2011jy,Baker:2012zs} which try to keep this parameterized framework as general as possible. Since in this work we focus on the Hu-Sawicki PPF formalism, in what follows we will call it PPF for short.

The current constraints on general $f(R)$ models within the PPF formalism are $B_0<0.42(95\%{\rm C.L.})$ 
by using CMB and ISW-galaxy correlation, and $B_0<1.1\times10^{-3}$ at $95\%$ C.L. \cite{Lombriser:2010mp}
using a larger set of data, such as WMAP5 \cite{Dunkley:2008ie}, Arcminute Cosmology Bolometer Array Receiver (ACBAR) \cite{Kuo:2006ya}, Cosmic Background Imager (CBI) \cite{Readhead:2004gy}, Very Small Array (VSA) \cite{Grainge:2002da}, Supernova Cosmology Project (SCP) Union \cite{Kowalski:2008ez}, Supernovae and $H_0$ for the Equation of States (SHOES) \cite{Riess:2009pu}, BAO data \cite{Percival:2009xn}, ISW-galaxy correlation ISWWLL likelihood code \cite{Ho:2008bz,Hirata:2008cb}, the $E_G$ measurement, probing the relation between weak gravitational lensing and galaxy flows \cite{Reyes:2010tr}, cluster abundance (CA) from the likelihood code of \cite{Mandelbaum:2012ay}, as well as the $\sigma_8$ measurement of the Chandra Cluster Cosmology Project (CCCP) \cite{Vikhlinin:2008ym}. The main constraint comes from cluster abundance.
The first constraint on $f(R)$ models from local cluster abundance from the several Gpc scale to the tens of Mpc scale is derived in \cite{Schmidt:2009am}, and the new constraint $|f_{R0}|<3.5\times10^{-3}$
at 1D marginalized 95\% confidence level through cluster-galaxy lensing of maxBCG clusters at $(0.2 − 20)$ Mpc scale from the Sloan Digital Sky Survey (SDSS) \cite{ Koester:2007bg,Abazajian:2008wr}, is obtain in \cite{Lombriser:2011zw}.
Besides what mentioned above, there exist many other parameterizations \cite{Bean:2010zq,Bertacca:2011wu,Linder:2005in,Baker:2011jy}. 

On the other hand, from the theoretical point of view, it has become clear that the cross-correlation between the Integrated Sachs-Wolfe (ISW) effect \cite{Sachs:1967er} and the gravitational Weak Lensing (WL) in CMB anisotropies can provide a wealth of cosmological information, including the dynamics of cosmic acceleration~\cite{Goldberg:1999xm, Verde:2002mu, Giovi:2003ri, Acquaviva:2004fv, DiValentino:2012yg}. Recently it has been shown that this effect will provide the most important contamination to the local-type primordial bispectrum \cite{Smith:2006ud,Hanson:2009kg, Mangilli:2009dr,Lewis:2011fk,Lewis:2012tc,Pearson:2012ba,Junk:2012qt}. In order to detect the primordial non-Gaussian signals, one needs to accurately subtract this secondary anisotropy in the {\it Planck} data. In fact various forecasts indicate that the {\it Planck} mission can measure this secondary effect with a high precision~\cite{Smith:2006ud,Hanson:2009kg, Mangilli:2009dr, Lewis:2011fk,Lewis:2012tc,Pearson:2012ba,Junk:2012qt}. 
From the observational point of view, determining the cosmological parameters will require not only additional measurements, but also characterizations of the data beyond the power spectrum. ISW-Lensing bispectrum can represent one of these novel statistical variables. In this paper we focus on constraining modified gravity models via future {\it Planck} ISW-Lensing bispectrum data.

\section{Parameterization of Modified Gravity}
Compared with the PPN formalism, PPF is a parameterization of a series of differential equations, rather than a specific static/stationary solution.
This is because what we are interested in cosmology is not only inhomogeneous but also time-varying quantities.
Fortunately, in some dominant regime for measures of the rate of structure growth, such as the weak lensing and peculiar velocity surveys, the time derivatives of perturbations can be neglected relative to spatial derivatives. This approximation leads to the so called quasi-static approximation, which can reduce the time differential equations into simple algebra relations. However, on the super-horizon scales the long wavelength perturbations have enough time to feel the changes of the gravitational potentials, consequently, the time derivative terms in the differential equations cannot be neglected. One explicit example is the ISW effect. In this section, we first recall one of the conventional PPF parametrization which is mainly inspired by the quasi-static dynamics, the Bertschinger-Zukin parameterization~\cite{Bertschinger:2008zb}; and then we introduce a more suitable parameterization for our purposes, the Hu-Sawicki PPF formalism, which properly takes the super-horizon perturbation evolution into account, even for large value of $B_0$ ($f_{R0}$).

\subsection{Bertschinger-Zukin parameterization}
As a straightforward extension of the PPN formalism, Bertschinger-Zukin formalism (BZ) \cite{Bertschinger:2008zb} proposed a quite efficient parameterization of modified gravity in the quasi-static regime, where the dynamical behavior of the temporal and spatial gravitational potentials are governed by two modified Poisson-like equations
\bea
-k^2\Phi&=&4\pi G_{\Phi}(t,\bk) a^2\rho_m\d\;,\\
-k^2\Psi&=&4\pi G_{\Psi}(t,\bk) a^2\rho_m\d\;,\eea
with
$G_{\Psi}(t,\bk)=\g(t,\bk) G_{\Phi}(t,\bk)$\footnote{
Notice that the metric convention of Bertschinger-Zukin is different w.r.t. Hu-Sawicki, we have:
$\Phi_{{\rm BZ}}=\Psi_{{\rm HS}}$ and $\Psi_{{\rm BZ}}=-\Phi_{{\rm HS}}$. In this paper, except in this subsection, we follow Hu-Sawicki conventions.}. Given the above observation, Bertschinger and Zukin proposed the following parameterization for $f(R)$ model
\be\label{BZPara}
\f{G_{\Phi}}{G_0}=\f{1+\a_1k^2a^s}{1+\a_2k^2a^s}\;,\quad \g=\f{1+\b_1k^2a^s}{1+\b_2k^2a^s}\;,\ee
with
\be\label{QSApp}
\Psi(t,\bk) = \gamma(t,\bk)\Phi(t,\bk)\, .
\ee
where $G_0$ is the Newton constant at the present epoch.
We can easily see that in the quasi-static approximation time derivative terms are neglected so that longwavelength perturbations cannot be properly taken into account (e.g., $\gamma \rightarrow 1$ for $k \rightarrow 0$, while in general, modified gravity theories can show distinctive features also on the longwavelength regime)\footnote{In \cite{Bertschinger:2008zb} authors also discussed the super-horizon perturbation dynamics which, however, is not captured by parameterized equation (\ref{BZPara}).} However, authors of \cite{Hojjati:2012rf} recently demonstrated that for small values of $B_0$ ($f_{R0}$), 
the parameterization of \cite{Bertschinger:2008zb} is practically good enough for the current data analysis purpose. 
In our following studies, we also investigate this kind of parameterization. Constraining modified gravity models via the ISW-Lensing bispectrum forces to take the ISW effect into account properly. Given this request, in what follows we turn to Hu-Sawicki PPF formalism in that it allows to join the superhorizon to the subhorizon regime. 

\subsection{\label{IIB}Hu-Sawicki model in PPF formula}
The PPF parameterization is defined by 3 functions $g(\ln a,k_H), f_{\zeta}(\ln a), f_G(\ln a)$ and 1 parameter $c_{\Gamma}$. They correspond to the metric ratio, the super-horizon relationship between the metric and density, the deviation of Newton constant on super-horizon scale from that on quasi-static regime, and the relationship between the transition scale and the Hubble scale~{\cite{Hu:2007pj}.
The 4th order nature of $f(R)$ theories provides enough freedom to reproduce any cosmological background history by an appropriate choice of the $f(R)$ function \cite{Song:2006ej}. For simplicity, in this paper we concentrate on one specific parameterization form, namely Hu-Sawicki model~\cite{Hu:2007nk}, which can satisfy the local solar system tests 
\be\label{Hu-Sawicki}
f(R)=-m^2\f{c_1(R/m^2)^n}{c_2(R/m^2)^n+1}\;,\ee
where the mass scale reads 
\be\label{MassScale}
m^2\equiv H_0^2\Omega_m=(8315{\rm Mpc})^{-2}\le(\f{\Omega_m h^2}{0.13}\ri)\;.\ee
The non-linear terms in $f(R)$ introduce higher-order derivatives into this theory. However, we are more familiar with a second-order derivative theory. Fortunately, we can reduce the derivatives to second order by
defining an extra scalar field $\chi\equiv\dd f/\dd R$, namely the ``scalaron'', which absorbs the higher derivatives.   
The square of the Compton wavelength of the scalaron in units of the Hubble length 
is defined as
\be\label{comp_wav}
B=\f{f_{RR}}{1+f_R}R'\f{H}{H'}\;,\ee
with $f_R=\dd f/\dd R$, $f_{RR}=\dd^2 f/\dd R^2$ and ${~}'\equiv\dd /\dd \ln a$.

At the high curvature regime, (\ref{MassScale}) can be expanded w.r.t. $m^2/R$ as 
\be\label{Hu-SawickiApp}
\lim_{m^2/R\>0}f(R)\approx -\f{c_1}{c_2}m^2+\f{c_1}{c_2^2}m^2\le(\f{m^2}{R}\ri)^n+\cdots\;.\ee
From Eq. (\ref{Hu-SawickiApp}) we can see that, the first and second terms represent a cosmological term and a deviation from it, respectively.  In order to mimic a $\Lambda$CDM evolution on the background, the vale of $c_1/c_2$ can be fixed as~\cite{Hu:2007nk}
\be\label{C1OverC2}
\f{c_1}{c_2}\approx 6\f{\Omega_{\Lambda}}{\Omega_m}\;.\ee
By using the relation (\ref{C1OverC2}) the number of free parameters can be reduced to two. 
From the above analysis, we can see that, strictly speaking, due to the appearances of correction terms to the cosmological constant, Hu-Sawicki model cannot exactly mimic $\Lambda$CDM. Since $m^2/R$ increase very fast with time, the largest value (at the present epoch) is $m^2/R\sim0.03$, the largest deviation to $\Lambda$CDM background happens in the case $n=1$ with $1\%$ errors, corresponding to $m^2/Rc_2\sim 0.01$ in Eq.(\ref{Hu-SawickiApp}). For large $n$ values, such as $n=4,6$ we can safely neglect this kind of theoretical errors. We have checked that for $n=1$ case, this $1\%$ deviation from $\Lambda$CDM will bring us a $10\%$ errors in the variance of the parameter $B_0$, while for $n=4,6$ our results are not affected. 

Without loss of generality we can choose the two free parameters to be ($n,c_2$). However, as shown in \cite{Song:2006ej}, for more general $f(R)$ models the $\Lambda$CDM evolution on the background can be reproduced exactly by only introducing one free parameter. This means that there exists some degeneracy between the two parameters,  as shown in Fig. \ref{fig:c2}. Usually General Relativity (GR) is recovered when 
\be\label{GRcondition}
B_0=0\;.\ee
Then, from Fig. \ref{fig:c2} we can see no matter what value $n$ takes, we are always allowed to set 
$B_0=0$ by adjusting $c_2$. Furthermore, in order to mimic $\Lambda$CDM on the background, $c_2$ and $n$ need to satisfy one constraint: the first term in the denominator of (\ref{Hu-Sawicki}) should be much larger than the second. The blank area in Fig. \ref{fig:c2} corresponds those parameter regions that are ruled out by $c_2(R/m^2)^n>1$. Actually, this condition gives
\begin{subnumcases}{B_0^{{\rm max}}=}
  0.1\;,\quad (n=1)\;,\nn\\
  1.2\;,\quad (n=4)\;,\label{B0Max}\\
  4.0\;,\quad (n=6)\;.\nn
\end{subnumcases}
In the approximation (\ref{Hu-SawickiApp}), we assume $m^2/R\>0$, however, with the expansion of the  universe this approximation starts to be not so accurate for our purposes. In order to obtain our results as accurate as possible, in the following calculation we use the exact expression (\ref{Hu-Sawicki}) in place of the approximated one.  
\begin{figure}[h]
\begin{center}
  \includegraphics[width=0.35\textwidth]{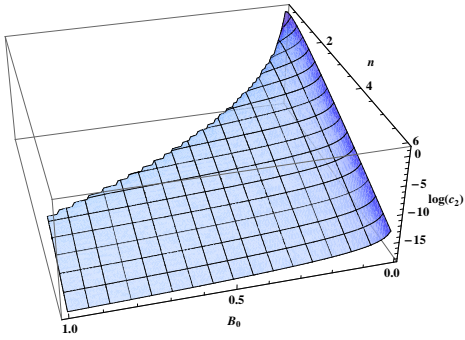}
  \caption{\label{fig:c2}The degeneracy between parameters $c_2$ and $n$.}
\end{center}
\end{figure}

We have decided to study the Hu-Sawicki model within the PPF formalism for the particular relevance of this model, and to exploit the generality of the PPF formalism for the study of the perturbations, using for the computations the publicy available PPF module \cite{Fang:2008sn} of CAMB~\cite{Lewis:1999bs}. In Ref.~\cite{next} we apply the same formalism to more general $f(R)$ models and other models of modified gravity. 
As stated previously, the idea of the Hu-Sawicki PPF formalism is to parameterize gravity theory
in order to join different scale regimes. 
In what follows, we will briefly review 
this parameterization.  
Firstly, we define the metric ratio as
\be\label{metric_ratio}
g=\f{\Phi+\Psi}{\Phi-\Psi}\;.\ee
In principle, $g$ is completely equivalent to the function $\gamma(t,k)$ introduced by 
Bertschinger and Zukin in \cite{Bertschinger:2008zb}.

\subsubsection{Super-Horizon Regime}
On super-horizon scales, the dynamical equation of motion for the spatial curvature 
potential has the following form~\cite{Hu:2007pj}
\bea\label{SHeq}
&&\Phi''+\le(1-\f{H''}{H'}+\f{B'}{1-B}+B\f{H'}{H}\ri)\Phi'\nn\\
&&+\le(\f{H'}{H}-\f{H''}{H'}+\f{B'}{1-B}\ri)\Phi=0\;,\quad (k_H\>0)\;.\eea
Furthermore, the
comoving curvature conservation relation ($\zeta'=0$) gives
\be\label{SHeq2}
\Psi=\f{-\Phi-B\Phi'}{1-B}\;,\quad (k_H\>0).\ee 
On super-horizon scales, because the spatial gradients of the perturbations can be neglected compared with their temporal evolution, the metric ratio reduces to a function $g_{{\rm SH}}(\ln a)$ dependent only on time
\be\label{gSH}
g(\ln a,k_H=0)=g_{{\rm SH}}(\ln a)=\f{\Phi+\Psi}{\Phi-\Psi}\;.\ee
The relevant phenomenon about the decay/growth of the gravitational potentials
in the CMB photons is the ISW effect. 
In Fig. \ref{fig:cosvar} we compute the ISW effect in the modified gravity models we consider in this paper. 
Our calculations demonstrate that ISW effect is quite sensitive to the value of the parameter $n$ in the Hu-Sawicki model.
The left plot shows that, for the Hu-Sawicki model with $n=1$, the ISW effect is always decreasing with increasing of $B_0$. 
However, for $n=4$ case, the ISW effect firstly decreases monotonically then bounce again after reaching some critical values. This behaviors are also reflected on our likelihood analysis (see later, Fig. \ref{fig:like}).
\begin{figure}[h]
\begin{center}
  \includegraphics[width=0.35\textwidth]{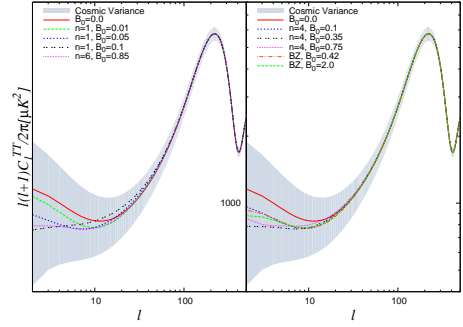}
  \caption{\label{fig:cosvar}
ISW effect in modified gravity models: Bertschinger-Zukin (BZ) model, Eq.(\ref{BZPara}) and $f(R)$ Hu-Sawicki model, Eq.(\ref{Hu-Sawicki}).}
\end{center}
\end{figure}

\subsubsection{Quasi-Static Regime}
In the quasi-static regime, all the time derivative terms are neglected w.r.t. 
the spatial gradients. At the linear perturbation order, this brings a lot of  
simplifications, because we can reduce the ordinary differential equations into 
algebraic relations. 
As an example, 
the modified Poisson equation takes the following form~\cite{Hu:2007pj}
\be\label{PoissonEq}
k^2\Phi_-=\f{4\pi G}{1+f_G}a^2\rho_m\Delta_m\;,\ee
where $f_G$ represents the deviation from the standard gravitational constant (here $\Phi_-=(\Phi-\Psi)/2$). 
Furthermore, for $f(R)$ models, we have 
\be\label{f_G}
f_G=f_R\;,\quad 
g\>g_{{\rm QS}}=-1/3\;,\ee
below the Compton wavelength.

\subsubsection{Matching the superhorizon and the subhorizon regimes}
In order to match the super-horizon scale behavior, an additional term $\Gamma$
must be introduced to the modified Poisson equation~\cite{Hu:2007pj}
\be\label{InterEq}
k^2\Big[\Phi_-+\Gamma\Big]=4\pi Ga^2\rho_m\Delta_m\;,\ee
where $\Gamma$ satisfy the following equation
\be\label{GammaEq}
(1+c_{\Gamma}^2k_H^2)\Big[\Gamma'+\Gamma+c_{\Gamma}^2k_H^2(\Gamma-f_{G}\Phi_-)\Big]=S\;,\ee
with the source term 
\bea\label{source}
S&=&-\le[\f{1}{g+1}\f{H'}{H}+\f{3}{2}\f{H_m^2}{H^2a^3}(1+f_{\zeta})\ri]\f{V_m}{k_H}\nn\\
&&+\le[\f{g'-2g}{g+1}\ri]\Phi_-\;.\eea
In Eq.(\ref{GammaEq}), the coefficient $c_{\Gamma}$ represents the relationship between
the transition scale and the Hubble scale, and the function $f_{\zeta}$ the relationship between
the metric and the density perturbation. For $f(R)$ models, we have
\be\label{c_GammaAndf_zeta}
c_{\Gamma}=1\;,\quad f_{\zeta}=c_{\zeta}g\;,
\quad c_\zeta\approx1/3\;.\ee
Up to now, the only unknown function in the intermeadiate regime is 
the metric ratio. 
An interpolating function is proposed in \cite{Hu:2007pj} as
\be\label{g_interplt}
g(\ln a,k)=\f{g_{{\rm SH}}+g_{{\rm QS}}(c_gk_H)^{n_g}}{1+(c_gk_H)^{n_g}}\;,\ee
with
\be\label{HuCoeff}
g_{{\rm QS}}=-1/3\;,\quad n_g=2\;,\quad c_g=0.71\sqrt{B(t)}\;.\ee

\section{\label{Sec:ISW-L} ISW-Lensing Bispectrum}
The primordial local non-Gaussianity is generated when the short-wavelength fluctuations are modulated 
by the long-wavelength modes which cross the horizon much earlier than the former. This kind of interactions will 
induce some correlations between the random fluctuations on large scales and those on small scales. 
In principle, any mechanisms through which small and large scale fluctuations couple together will generate 
similar features as those in the primordial local-type non-Gaussianity.

The late-time cross-correlation between ISW and 
Weak Lensing provides one such mechanism.
On one hand, when the universe starts to accelerate at late-time, the decay of the gravitational potential generates the 
secondary anisotropies in the CMB anisotropies, the ISW effect 
\be\label{ISW}
\f{\d T}{T}(\hat{{\bf n}})|_{{\rm ISW}}=\int\dd \chi (\Phi-\Psi)_{,\tau}(\hat{{\bf n}},\chi)\;,\ee
where $\tau$ and $\chi$ are the conformal time and comoving distance, respectively.
On the other hand, the gravitational lensing induced by the fluctuations of matter density deflects 
the paths of CMB photons as they travel from the last-scattering surface to the
observer
\bea\label{WL}
\d \tld T(\hat{{\bf n}})&=&\d T(\hat{{\bf n}}+\boldsymbol\6\phi)\nn\\
&\simeq &\d T(\hat{{\bf n}})+\Big[(\boldsymbol{\6}\phi)\cdot(\boldsymbol{\6}\d T)\Big](\hat{{\bf n}})\;,\eea
where the lensing potential is defined as
\be\label{GravPoten}
\phi(\hat{\bf n})=-\int_0^{\chi_{\ast}}\dd \chi\f{\chi_{\ast}-\chi}{\chi_{\ast}\chi}\Big(\Phi-\Psi\Big)(\hat{\bf n},\chi)\;.\ee
From Eq.(\ref{ISW}) and (\ref{WL}), we can see that Weyl potential ($\Phi-\Psi$) sources both ISW and WL, so the long-wavelength mode from ISW couples with the short-wavelength mode from WL. Consequently, this kind of correlation will induce the secondary ISW-Lensing bispectrum, which has been recognized as the most dominant contamination of local-type primordial bispectrum~\cite{Smith:2006ud,Hanson:2009kg,Mangilli:2009dr, Lewis:2011fk,Lewis:2012tc,Pearson:2012ba}

Going to harmonic space and computing the reduced bispectrum, one can obtain~\cite{Goldberg:1999xm,Verde:2002mu,Hanson:2009kg,Komatsu:2010hc} 
\bea\label{reducebi}
b_{l_1l_2l_3}^{{\rm ISW-L}}&=&\le[\f{-l_1(l_1+1)+l_2(l_2+1)+l_3(l_3+1)}{2}C^{{\rm T}}_{l_2}C^{\phi T}_{l_3}\ri.\nn\\
&&+5{\rm perm.}\Big]\;.\eea
where $C^{\phi T}_l=\langle  \phi^*_{lm}  a^{\rm ISW}_{lm} \rangle$. 
Since the gravitational lensing is just remapping the primordial CMB temperature and polarization fields via the gradient of a foreground potential, from the above expression we can see that ISW-Lensing bispectrum includes information from both the primordial fluctuations and the secondary ones. This fact is important for the understanding of the spectrum-bispectrum joint analysis, because the bispectrum can correlate with the power spectrum through both  the temperature-temperature mode of power spectrum ($C^{{\rm T}}_l$)  and the secondary lensing-temperature cross power spectrum ($C^{\phi T}_l$), as detailed in Sec.\ref{sec.joint}.
\begin{figure}[h]
\begin{center}
  \includegraphics[width=0.45\textwidth]{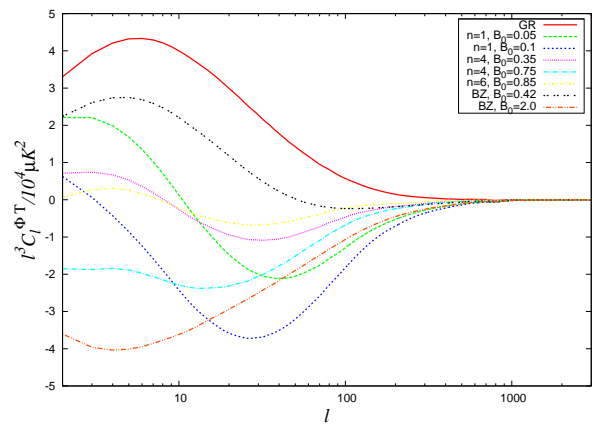}
  \caption{\label{fig:C_PhiTemp}The cross spectrum between lensing potential and temperature $l^3C_l^{\Phi T}$. }
\end{center}
\end{figure}

\begin{figure}[h]
\begin{center}
  \includegraphics[width=0.45\textwidth]{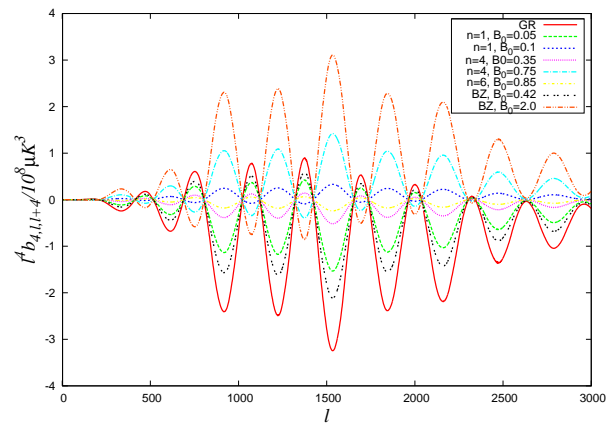}
  \caption{\label{fig:reduced}The squeezed slice of reduced bispectrum $l^4b_{4,l,l+4}$. }
\end{center}
\end{figure}
In Fig.\ref{fig:C_PhiTemp}, we plot the lensing-temperature cross power spectrum for several different 
gravity models. From this figure, we can explicitly see the non-linear dependence of $C^{\phi T}_l$ on $B_0$:
firstly, in the GR limit $B_0\>0$ (red solid curve), the cross-correlation is large and positive; secondly, with 
increasing of $B_0$, the cross-correlation decreases and oscillates around zero (green dotted curve for $n=1$, 
pink dashed curve for $n=4$ Hu-Sawicki models, and black dotted-dashed curve for Bertschinger-Zukin model); finally, with further increasing of $B_0$, the cross-correlation amplitude increase again but negatively (dotted deep blue curve for $n=1$, dashed-dotted light blue curve for $n=4$ Hu-Sawicki models, and dashed-dotted orange curve for the Bertschinger-Zukin model). The reasons of these phenomena lie in the non-linear dependence of ISW effect on $B_0$, i.e.
the first few multipoles firstly decrease and then bounce with the increasing of $B_0$, see Fig.\ref{fig:cosvar}.
The other important feature is that the cross-correlations are suppressed greatly after the first hundred multipoles, it almost vanishes in the regime $l\geq 1000$,  because the cross-correlation is driven by the late time ISW effect.

We also investigate the squeezed profile of reduced ISW-Lensing bispectrum, see Fig.\ref{fig:reduced}. 
From this figure, we can easily see the modulation of short-wavelength mode by the long-wavelength mode.

\section{Forecasts and results}
In this section, we will present the joint likelihood analysis of the CMB power spectrum and ISW-Lensing bispectrum for several modified gravity models by using the specifications of WMAP and Planck experiments. 
In this paper, we implement the Hu-Sawicki model by using the PPF module \cite{Fang:2008sn} of CAMB \cite{Lewis:1999bs} and the Bertschinger-Zukin model by using MGCAMB \cite{Hojjati:2011ix}.
Our results show that the constraint on $B_0$ is quite model dependent. However, generally speaking, the ISW-Lensing bispectrum can constrain cosmological parameters at the same level as the power spectrum does. Consequently, the joint analysis of the power spectrum and the ISW-lensing bispectrum can improve the $B_0$ error bars by a factor $1.14$ to $5.32$.


\subsection{Power spectrum}
In this subsection, we forecast constraints on $B_0$ from observations
of the CMB temperature power spectrum. To this aim we firstly consider
the simple likelihood extracted from the chi-squared
variable:
\be\label{chi_P}
\chi^2_{{\rm P}}=\sum_{l=2}^{3000}\f{\Big[C_l^{th}-C_l^{fid}\Big]^2}{\sigma_{{\rm P},l}^2}\;,\ee
where $C_l$ is the lensed "TT" power spectrum, and the variance is 
\be\label{Var_P}
\sigma_{{\rm P},l}^2=\f{2}{(2l+1)}\Big(b^2_l C^{T}_l+N_l\Big)^2\;,\ee
$N_l$ and $b_l$ denoting the experimental noise and beam
respectively (listed in Tab.\ref{tab:experiment}).
We allow only $B_0$ to vary while fixing all other 
cosmological parameters to their WMAP7yr best-fit values, as in
\cite{DiValentino:2012yg}. 
The $B_0$-likelihood obtained in this way is shown in figure 
Fig.\ref{fig:like} and \ref{fig:like2}. 
In agreement with previous studies
\cite{Lombriser:2010mp,DiValentino:2012yg}, we find that its behaviour
is highly non-Gaussian and for this reason we cannot rely on a
standard Fisher matrix approach in order to forecast error bars on
$B_0$. Despite this issue we still computed a full 7-parameter Fisher
matrix as a mean to get an order of magnitude estimate of the
correlation between $B_0$ and standard $\Lambda CDM$ parameters. Our
results are summarized in detail in Appendix \ref{App:Fisher} and show
that there are no significant degeneracies between $B_0$ and other
parameters  (the largest cross-correlation coefficient we find is
$\Omega_b h^2$-$B_0=-0.18$). 
This in turn justifies the simple 1-parameter likelihood approach used in this
section in order to study $B_0$ error bars.

\begin{table}
\caption{\label{tab:like1}$B_0$ error bars ($68\%$ CL) from power spectrum $\chi^2$ for the Hu-Sawicki model}
\begin{ruledtabular}
\begin{tabular}{cccccccc}
$n$ & WMAP & PLANCK \\
\hline
$1$		&	$4.20\times10^{-2}$	&	$6.68\times10^{-5}$ \\
$4$		&	-------- &	$2.06\times10^{-3}$ \\
$6$     	&   -------- &   	$1.51\times10^{-2}$ \\
\end{tabular}
\end{ruledtabular}
\footnotetext{The blank elements in the table represent that WMAP data cannot efficiently constrain $B_0$ parameter in Hu-Sawicki model with $n=4,6$, because the likelihood does not vanish in the tails which can be read from Fig.\ref{fig:like} and \ref{fig:like2}.}
\end{table}

\begin{table}
\caption{\label{tab:like2}$B_0$ error bars ($68\%$ CL) from power spectrum $\chi^2$ for the Bertschinger-Zukin model}
\begin{ruledtabular}
\begin{tabular}{cccccccc}
Ideal ($l_{{\rm max}}=1000$) & PLANCK \\
\hline
$1.63\times10^{-1}$	&	$1.10\times10^{-2}$ \\
\end{tabular}
\end{ruledtabular}
\footnotetext{In order to compare with \cite{DiValentino:2012yg}, we forecast $B_0$ error bars in the ideal experiment with $l_{{\rm max}}=1000$ instead of WMAP.}
\end{table}

\begin{figure}[h]
\begin{center}
  \includegraphics[width=0.45\textwidth]{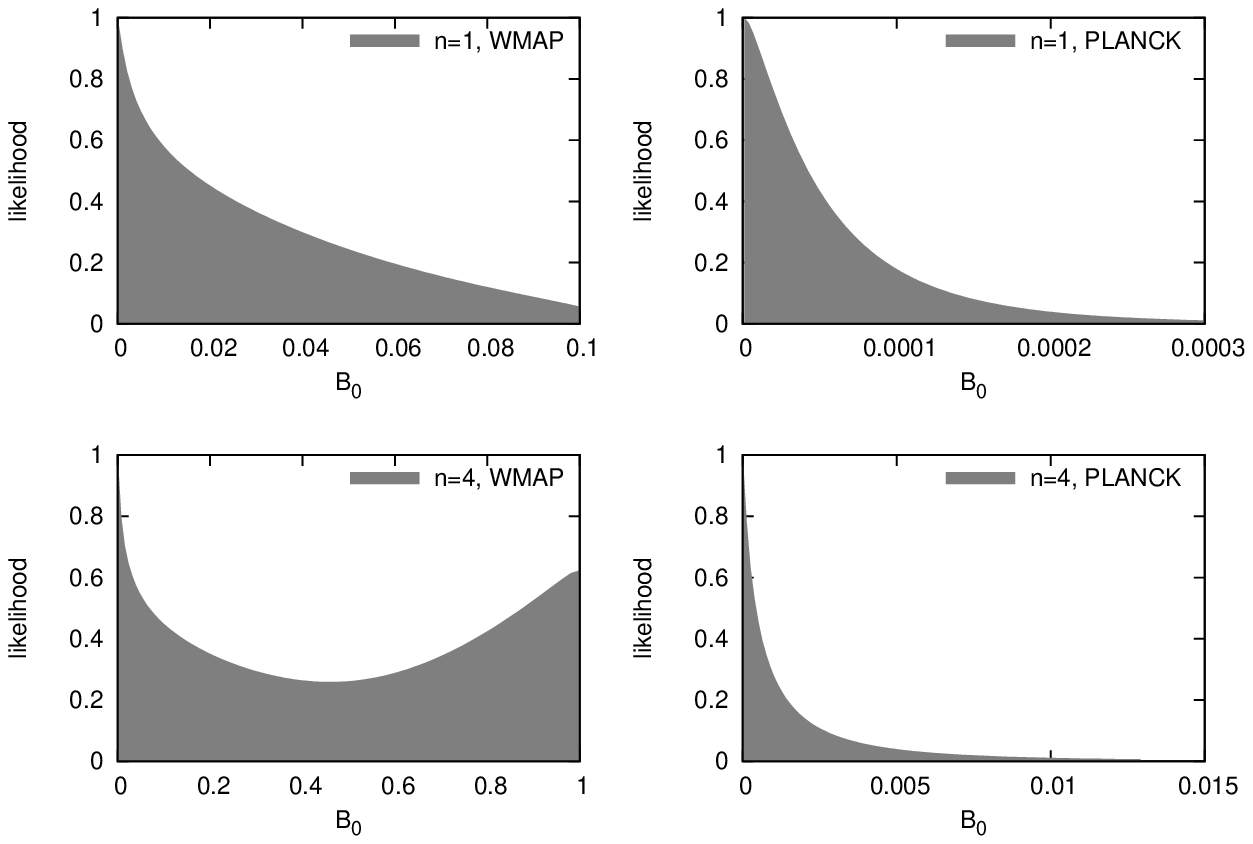}
  \caption{\label{fig:like}WMAP/PLANCK experiments power spectrum likelihood of $B_0$, for $n=1$ (top) and $n=4$ (bottom).}
\end{center}
\end{figure}

\begin{figure}[h]
\begin{center}
  \includegraphics[width=0.45\textwidth]{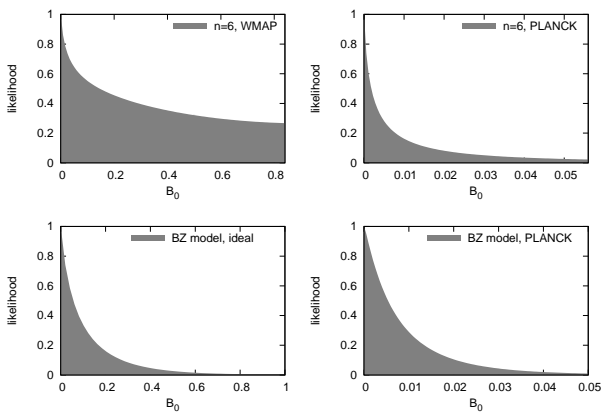}
  \caption{\label{fig:like2}Hu-Sawicki model (n=6) and Bertschinger-Zukin model power spectrum likelihood.}
\end{center}
\end{figure}

\begin{figure}[h]
\begin{center}
  \includegraphics[width=0.45\textwidth]{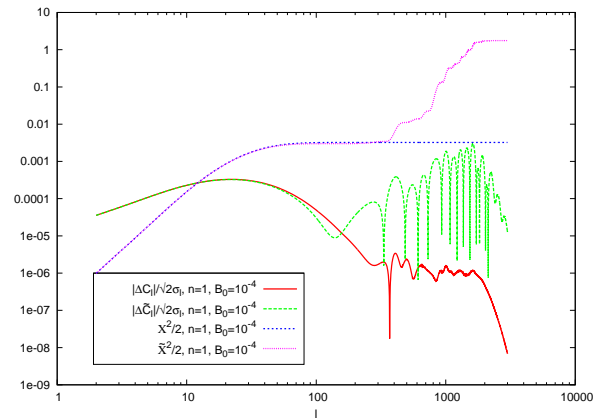}
  \caption{\label{fig:ratio3}. The demonstration of main signal of constraining $B_0$ in Hu-Sawicki model with $n=1, B_0=10^{-4}$. The red (unlensed) and green (lensed) curves shows the fractional difference in TT spectrum between $f(R)$ model and general relativity. And the blue (unlensed) and pink (lensed) curves shows the increase of $\chi^2/2$ with respect to the maximum of multipoles.}
\end{center}
\end{figure}

We investigate the likelihood function for both WMAP and Planck experiments in three Hu-Sawicki models with different values of $n=1,4,6$ in Fig.\ref{fig:like} and \ref{fig:like2}. 
Furthermore, in order to
compare with the results in \cite{DiValentino:2012yg} we also include
the Bersthinger-Zukin model in Fig.\ref{fig:like2}. The corresponding
$68\%$ CL's are summarized in Tab.\ref{tab:like1} and
\ref{tab:like2}. 
Notice that the likelihood of Hu-Sawicki model for $n=4,6$ in Fig. \ref{fig:like} and Fig. \ref{fig:like2} shows that WMAP data does not have enough constraining power on parameter $B_0$. This is because with WMAP resolution we are insensitive to the lensing signal which carries most of late-time evolution information.
From the above results, we firstly see that
\emph{Planck} is able to reduce the error bar at least one order of
magnitude.
In Fig.\ref{fig:ratio3}, we demonstrate that these improvements mainly come from lensing signal. The $\chi^2$ for lensed TT spectrum (pink curve) increases very fast on small scales and saturate around multipoles $l\sim 2000$, while the unlenesd one (blue curve) only receives the contributions from late time ISW effect on large scales. In  Fig.\ref{fig:ratio3} we also plot the ratio $\Delta C_l/\sqrt{2}\sigma_l$ for the Hu-Sawicki model with parameter values $n=1, B_0=10^{-4}$, where $\Delta C_l$ is defined as
\begin{equation}
\Delta C_l\equiv C^T_l(B_0=10^{-4}) - C^T_l(B_0=0)\;,
\end{equation} 
and the variance $\sigma_l$ can be found in Eq. (\ref{Var_P}). 
From Fig.\ref{fig:ratio3} we can see that the ISW effect contribution to the final signal to noise ratio is not negligible in amplitude compared with the lensing part. However, the number of multipoles which ISW could change is much less than those of lensing effect.
Furthermore, we can also see that the $\chi^2$ for the lensed case (pink curve) saturates at multipoles ($l\sim 2000$) around $\mathcal O(1)$ for $B_0=10^{-4}$, which agrees with our global analysis.
We have checked that the other models considered in this paper share a similar behavior.

Notice that for the Bertschinger-Zukin model, our error bars
 for an ideal experiment (no noise, $l_{{\rm max}}=1000$) compared with those in~\cite{DiValentino:2012yg} ($\s=0.61$) are improved by a factor $3.74$. This is due to the fact that in our analysis we account for the weak lensing in the power spectrum.

\subsection{Bispectrum}
As already mentioned in the previous section, even for perfectly Gaussian
primordial perturbations ISW and WL effects generate some non-Gaussian features in the CMB.
In this subsection, we explore the possibility to constrain
cosmological parameters via the ISW-lensing bispectrum.
For a rotation-invariant sky, all the cosmological information is
encoded in the {\em angular-averaged bispectrum} (e.g.,
see~\cite{Komatsu:2001rj}), defined as:
\be\label{angle_mean_Bi}
B_{l_1l_2l_3} = \sqrt{\f{(2l_1+1)(2l_2+1)(2l_3+1)}{4\pi}}
\le(
 \begin{array}{ccc}
  l_1&l_2&l_3\\
  0&0&0\\
 \end{array}
  \ri)
 b_{l_1l_2l_3}\;,\ee 
where the {\em reduced bispectrum} $ b_{l_1l_2l_3}$ produced by the ISW-Lensing correlation is given by Eq. (\ref{reducebi}).
The averaged bispctrum is different from 0 only for $l_1, l_2, l_3$ triples that satisfy triangle inequality and parity conservation relationships:
\begin{subnumcases}{}
  l_1\leq l_2\leq l_3\;,\nn\\
  l_1+l_2+l_3={\rm even}\;,\\
  |l_j-l_k|\leq l_i\leq l_j+l_k\, , 
\end{subnumcases}
Following the approach of the previous section, we now consider the
chi-squared variable \cite{Goldberg:1999xm,Verde:2002mu,Komatsu:2001rj,Komatsu:2010hc}:
\be\label{chi_B}
\chi^2_{{\rm B}}=\sum_{l_1,l_2,l_3}
\f{\Big[B^{th}_{l_1l_2l_3}-B^{fid}_{l_1l_2l_3}\Big]^2}{\sigma_{{\rm B},l_1l_2l_3}^2}\;,\ee
where the bispectrum variance reads
\be\label{Var_B}
\sigma_{{\rm B},l_1l_2l_3}^2=\Delta_{l_1l_2l_3}\mathtt C^T_{l_1}\mathtt C^T_{l_2}\mathtt C^T_{l_3}\;,
\quad (\mathtt C^T_l= C^T_l+N_l)\;,\ee
with the weights
\begin{subnumcases}{\Delta_{l_1l_2l_3}=}
  6\;,\quad (l_1=l_2=l_3)\;,\nn\\
  2\;,\quad ({\rm two~~identical})\;,\\
  1\;,\quad ({\rm all~~different})\;.\nn
\end{subnumcases}

\begin{table}
\caption{\label{tab:like3}$B_0$ error bars ($68\%$ CL) from bispectrum $\chi^2$ for the Hu-Sawicki model}
\begin{ruledtabular}
\begin{tabular}{cccccccc}
$n$ & WMAP & PLANCK \\
\hline
$1$		&	-------- 	&	$2.90\times10^{-4}$ \\
$4$		&	-------- 	&	$1.89\times10^{-3}$ \\
$6$     &   --------     &   $6.29\times10^{-3}$ \\
\end{tabular}
\end{ruledtabular}
\footnotetext{The blank elements in the table represent that WMAP data cannot efficiently constrain $B_0$ parameter in Hu-Sawicki model, because the likelihood does not vanish in the tails
which can be read from Fig. \ref{fig:like3} and \ref{fig:like4}.}
\end{table}

\begin{table}
\caption{\label{tab:like4}$B_0$ error bars ($68\%$ CL) from bispectrum $\chi^2$ for the Bertschinger-Zukin model}
\begin{ruledtabular}
\begin{tabular}{cccccccc}
Ideal ($l_{{\rm max}}=1000$) & PLANCK \\
\hline
$1.92\times10^{-1}$	&	$3.85\times10^{-2}$ \\
\end{tabular}
\end{ruledtabular}
\footnotetext{We can see that the results from ideal experiment with $l_{{\rm max}}=1000$ is in agreement with those obtained in\cite{DiValentino:2012yg}.}
\end{table}

\begin{figure}[h]
\begin{center}
  \includegraphics[width=0.45\textwidth]{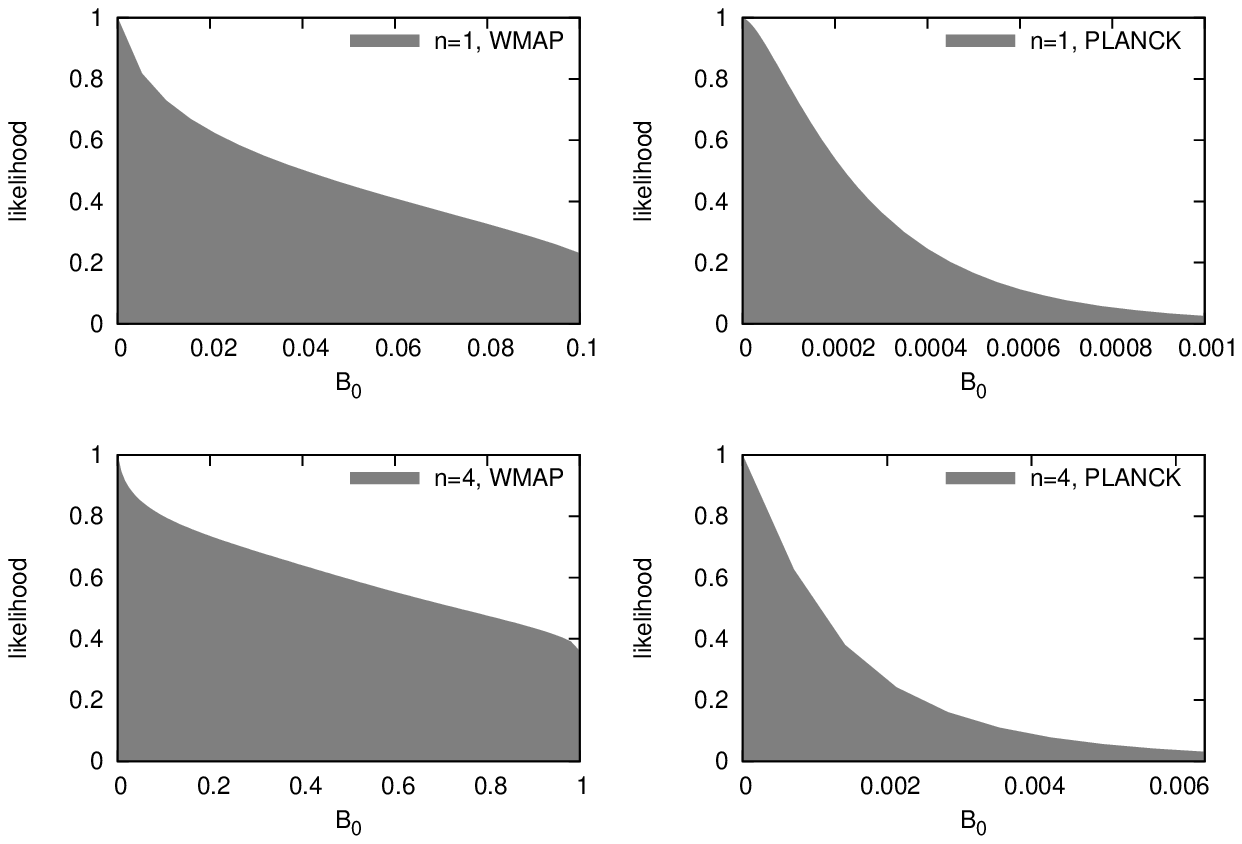}
  \caption{\label{fig:like3}WMAP/PLANCK experiments bispectrum likelihood of $B_0$, for $n=1$ (top) and $n=4$ (bottom).}
\end{center}
\end{figure}

\begin{figure}[h]
\begin{center}
  \includegraphics[width=0.45\textwidth]{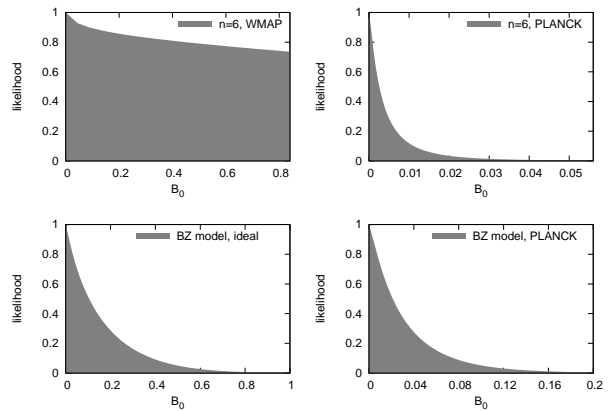}
  \caption{\label{fig:like4}Hu-Sawicki model (n=6) and Bertschinger-Zukin model bispectrum likelihood.}
\end{center}
\end{figure}
The CMB bispectrum likelihood for the Hu-Sawicki models with $n=1,4,6$ and the Bertschinger-Zukin model are plotted in Fig.\ref{fig:like3} and 
Fig.\ref{fig:like4}, respectively, and the corresponding $B_0$ error bars are listed in Tab.\ref{tab:like3}
and Tab.\ref{tab:like4}.
Similar as power spectrum case, WMAP data only cannot give any constraint on $B_0$.
Compared with the $B_0$ error bars obtained
from the power spectrum analysis, we do not find any significant
improvement from the bispectrum {\em alone}. In details, for the
Hu-Sawicki models with $n=4,6$ the ratio between the error bars from
the power spectrum and those from the bispectrum are $1.09$ and
$2.40$, respectively. For the Hu-Sawicki model with $n=1$ and the
Bertschinger-Zukin model, the error bars from the bispectrum are even
larger than those from the power spectrum. In the following section we
are however going to explore the possibility of improving our constraint
through a joint analysis of power spectrum and bispectrum.

\subsection{\label{sec.joint}Joint Analysis}
As already specified in the last part of Sec.\ref{Sec:ISW-L}, the power spectrum can correlate with the ISW-Lensing bispectrum via both the primordial and the secondary anisotropies (ISW and weak lensing). 
However, comparing with the primordial and ISW channels, the cross-correlation through weak lensing is sub-dominant. The reason is that cross-correlator of the power spectrum and bispectrum via weak lensing channel is an 8-point function of a Gaussian random fluctuations ($a_{lm}, \phi_{lm}$), while both the primordial and ISW are a 6-point function. Therefore, we will focus on the unlensed cross-correlation channel in what follows, i.e. the power spectrum cross-correlation with the bispectrum through unlensed fluctuations.

If we compose the power spectrum and the bispectrum into a vector $\Delta\equiv(\Delta C_l,\Delta B_{l_1l_2l_3})$, the joint likelihood function reads
\be\label{JointLikelihood}
\mathcal L \propto {\rm exp}\le(-\Delta\cdot{\rm CoV}^{-1}\cdot \Delta^{{\rm T}}\ri)\;,\ee
where the inverse of the covariance matrix can be written as a $2\times 2$ blocked form
\be\label{2by2Cov}
{\rm CoV}^{-1}\equiv\f{1}{\sigma^2_{{\rm P}}\sigma^2_{{\rm B}}-(\sigma^2_{{\rm PB}})^2}
\begin{bmatrix}
  \sigma^2_{{\rm B},l_1l_2l_3} & -\sigma^2_{{\rm PB},l_1l_2l_3l} \\
  -\sigma^2_{{\rm PB},ll_1l_2l_3} & \sigma^2_{{\rm P},l}
\end{bmatrix}\;.
\ee
When the power spectrum and the bispectrum are weakly correlated ($\sigma^2_{{\rm PB}}\ll \sigma^2_{{\rm P,B}}$), 
the above formula can be reduced into\cite{Sefusatti:2006pa}
\be\label{JointChi2}
\mathcal L \propto {\rm exp}(-\chi^2/2)\;,\quad 
\chi^2 = \chi^2_{{\rm P}}+\chi^2_{{\rm B}}+\chi^2_{{\rm PB}}\;.\ee

For the computation of cross likelihood we have 
\bea\label{chiPB}
&&\chi^2_{{\rm PB}} = \sum_{l,l_1,l_2,l_3}\f{-2\sigma^2_{{\rm PB},ll_1l_2l_3}\times\Delta C^{T}_l
\times\Delta B^{{\rm ISW-L}}_{l_1l_2l_3}}{\sigma^2_{{\rm P},l}\sigma^2_{{\rm B},l_1l_2l_3}}\;,\eea
where the variance reads
\bea\label{PBCross}
&&\sigma^2_{{\rm PB},ll_1l_2l_3}\equiv\Big\la\Delta\hat C^{T}_l\Delta\hat B^{{\rm ISW-L}}_{l_1l_2l_3}\Big\ra
=\nn\\
&&\sqrt{\f{(2l_1+1)(2l_2+1)(2l_3+1)}{4\pi}}
 \le(
 \begin{array}{ccc}
  l_1&l_2&l_3\\
  0&0&0\\
 \end{array}
 \ri)\nn\\
&& \times\le\{\half\Big[-l_1(l_1+1)+l_2(l_2+1)+l_3(l_3+1)\Big]\ri.\nn\\
&&\Big\la\hat C^{T}_l
\hat C^{\phi T}_{l_2}\hat C^{{\rm T}}_{l_3}\Big\ra + 5 {\rm perm.}\Big\}
-\bar C^{T}_l\bar B^{{\rm ISW-L}}_{l_1l_2l_3}\nn\\
&=&\sqrt{\f{(2l_1+1)(2l_2+1)(2l_3+1)}{4\pi}}
 \le(
 \begin{array}{ccc}
  l_1&l_2&l_3\\
  0&0&0\\
 \end{array}
 \ri)\nn\\
&& \times\le\{\half\Big[-l_1(l_1+1)+l_2(l_2+1)+l_3(l_3+1)\Big]\ri.\nn\\
&&\times\le[\f{2\d_{ll_3}}{(2l_3+1)}C^{{\rm T}}_lC^{\p T}_{l_2}C^{{\rm T}}_{l_3}
+\f{2\d_{ll_2}}{(2l_2+1)}C^T_lC^{\p T}_{l_2}C^{{\rm T}}_{l_3}
\ri.\nn\\
&&+\le.\le. \f{8\d_{ll_2}\d_{ll_3}}{(2l+1)^2}C^{{\rm T}}_lC^{\p T}_{l_2}C^{{\rm T}}_{l_3}\ri]
+5{\rm perm.}\ri\}\;,\eea
with 
\bea
\hat C^{\phi T}_l &=& \sum_{m,m'}\f{\delta_{mm'}\phi^{\ast}_{lm}a^{T}_{lm'}}{2l+1}\;,
\label{measureCl1}\\
\hat C^{{\rm T}}_l &=& \sum_{m,m'}\f{\delta_{mm'}a^{{\rm T}\ast}_{lm}a^{{\rm T}}_{lm'}}{2l+1}\;.
\label{meansureCl2}\eea
We emphasize again that here we cross-correlate ISW-Lensing bispectrum with \emph{unlensed} power spectrum.
From the above result, we can see that the first two terms in Eq.(\ref{PBCross}), which peak at the squeezed configurations, give the main contributions to the cross-correlation, while the last term only affect the equilateral configurations. Since our ISW-Lensing bispectrum signal peaks at the squeezed limit, the contributions from the third term is negligible. 
In Appendix.\ref{App:example}, we show a computation example about $\sigma^2_{\rm PB}$.


\begin{figure}[h]
\begin{center}
  \includegraphics[width=0.45\textwidth]{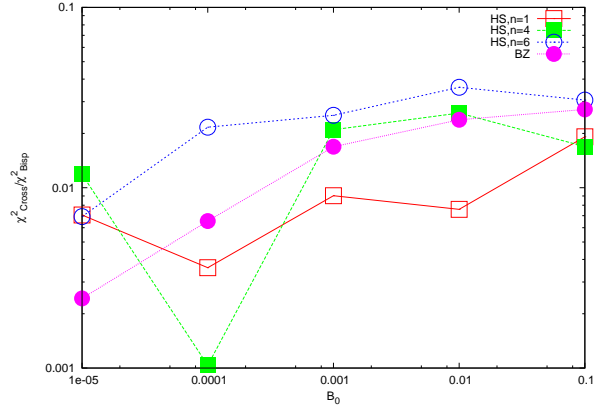}
  \caption{\label{fig:chi2}Likelihood ratio between cross-correlation and bispectrum $\chi^2_{\rm PB}/\chi^2_{\rm B}$. }
\end{center}
\end{figure}
In Fig.\ref{fig:chi2}, we show the likelihood ratio between cross-correlation and bispectrum for several $B_0$ values. 
This result indicates that the maximum correction from cross-correlations are $3\%$, so 
we can safely neglect the cross-correlations between the power spectrum and the bispectrum in the joint analysis. In Fig.\ref{fig:like5}, we plot the joint likelihoods of $B_0$ for the \emph{Planck} experiment, and the corresponding $68\%$ CL results are listed in Tab.\ref{tab:like5}. Comparing with the error bars from the power spectrum alone, there is an improvement of a factor ranging from $1.14$ to $5.32$ (depending 
on the specific modified gravity model).

\begin{figure}[h]
\begin{center}
  \includegraphics[width=0.45\textwidth]{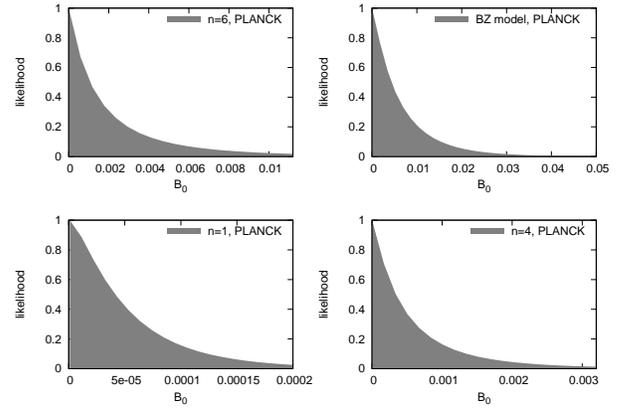}
  \caption{\label{fig:like5}Joint analysis results: spectrum and bispectrum product, no cross-correlation included.}
\end{center}
\end{figure}
\begin{table}
\caption{\label{tab:like5}$B_0$ error bars ($68\%$ CL) from the power spectrum-bispectrum joint analysis}
\begin{ruledtabular}
\begin{tabular}{cccccccc}
$n$/BZ & Joint & Power/Joint \\
\hline
$1$		&	$5.85\times10^{-5}$ & $1.14$\\
$4$		&	$6.90\times10^{-4}$ & $2.99$\\
$6$     &   $2.84\times10^{-3}$ & $5.32$\\
BZ		&   $7.87\times10^{-3}$ & $1.40$\\
\end{tabular}
\end{ruledtabular}
\end{table}

\section{conclusions}
The correlation between the Integrated Sachs Wolfe (ISW) effect and
gravitational lensing is expected to generate a bispectrum signature
in the CMB . In fact the \emph{Planck} satellite can detect this secondary non-Gaussianity~\cite{Smith:2006ud,Hanson:2009kg, Mangilli:2009dr, Lewis:2011fk,Lewis:2012tc,Pearson:2012ba}.
The high precision measurements of CMB anisotropies expected from
\emph{Planck} data will enable us to detect such signature at a high
level of significance ~\cite{Smith:2006ud,Hanson:2009kg,
  Mangilli:2009dr, Lewis:2011fk,Lewis:2012tc,Pearson:2012ba}. This
 will open the interesting possibility of actually using ISW-lensing
 bispectrum measurements as a tool to constrain cosmological
 parameters (see, e.g., ~\cite{Goldberg:1999xm, Verde:2002mu,
   Giovi:2003ri, Acquaviva:2004fv, DiValentino:2012yg}). Since the
 ISW-Lensing bispectrum is composed by the product of primordial power
 spectrum and the ISW-Lensing cross-spectrum, it carries information
 on both early (recombination) and late-time epochs. In this paper we investigated the 
possibility to use forthcoming \emph{Planck} data to constrain some
modified gravity models and their deviations from GR as a candidate to
explain the late-time 
acceleration of the universe, focusing in particular on the possibility to perform a joint analysis of CMB power spectrum and ISW-Lensing bispectrum.

Firstly we compared the ability to constrain cosmological parameters via CMB power spectrum only 
in WMAP and \emph{Planck} missions. Due to the high
resolution and sensitivity of the {\em Planck} dataset, \emph{Planck}
data will allow to improve the constraint on the modified gravity
parameter $B_0$ at least by one order magnitude compared to what one
can do with WMAP.
Secondly we performed a joint power spectrum/bispectrum analysis. 
Our results show that the joint
analysis could reduce the $B_0$ 1-$\sigma$ error bar by a factor
ranging from $1.14$ to $5.32$ w.r.t. to what can be achieved with the
power spectrum alone. The improvement factor depends on the
model considered: for the Hu-Sawicki models with large value of $n$
the improvements are quite significant unlike those with small value
of $n$ and the Bertschinger-Zukin model. Furthemore,  we have provided
the full formulae to compute the cross-correlation between the power
spectrum and the bispectrum (that are necessary in the joint-analysis
approach). 
One of our main results is that this cross-correlation can be safely neglected and this conclusion seems quite model independent. 


\begin{table}
\caption{\label{tab:summary}$B_0$ error bars ($68\%$ CL) summary for Planck experiment}
\begin{ruledtabular}
\begin{tabular}{cccccccc}
Model & Pow & Bisp & Joint \\
\hline
$1$		&	$6.68\times10^{-5}$	&	$2.90\times10^{-4}$	&	$5.85\times10^{-5}$	\\
$4$		&	$2.06\times10^{-3}$	&	$1.89\times10^{-3}$	&	$6.90\times10^{-4}$	\\
$6$     	&   	$1.51\times10^{-2}$ 	&   	$6.29\times10^{-3}$ 	&	$2.84\times10^{-3}$	\\
BZ		&	$1.10\times10^{-2}$ 	&	$3.85\times10^{-2}$	&	$7.87\times10^{-3}$	\\
\end{tabular}
\end{ruledtabular}
\end{table}

\begin{acknowledgments}
We we are greatly indebted to Wayne Hu, Alan Heavens, Daniele Bertacca for their helpful 
discussions and correspondences. This research has been partially supported by the ASI/INAF Agreement 
I/072/09/0 for the Planck LFI Activity of Phase E2 and by the PRIN 2009 
project "La Ricerca di non-Gaussianita` Primordiale"
\end{acknowledgments}

\appendix
\label{App:Fisher}
\section{\label{App:Fisher}CMB power spectrum Fisher matrix analysis}
In this Appendix, we present the results of our $C_l$ Fisher matrix analysis. 
The Fisher matrix for CMB anisotropies
and polarization is \cite{Seljak:1996ti,Zaldarriaga:1996xe,Kamionkowski:1996zd,Bond:1997wr,Zaldarriaga:1997ch,Eisenstein:1998hr}
\bea\label{Fisher1}
{\bf F}_{ij}=\sum_l\sum_{X,Y}\f{\6 C_{Xl}}{\6 p_i}\Big({\rm Cov}_l\Big)^{-1}_{XY}\f{\6 C_{Yl}}{\6 p_j}\;,\eea
where the covariance matrix is defined as 
\bea\label{FisherCov}
\Big({\rm Cov}_l\Big)_{{\rm TT}}&=&\f{2}{(2l+1)f_{{\rm sky}}}\Big[C_{Tl}+\bar w_T^{-1}\bar B_l^{-2}\Big]^2\;,\\
\Big({\rm Cov}_l\Big)_{{\rm EE}}&=&\f{2}{(2l+1)f_{{\rm sky}}}\Big[C_{El}+\bar w_P^{-1}\bar B_l^{-2}\Big]^2\;,\\
\Big({\rm Cov}_l\Big)_{{\rm TE}}&=&\f{2}{(2l+1)f_{{\rm sky}}}C^2_{Cl}\;,
\eea
with
\bea
&&w_{X,c}=\Big(\sigma_{X,c}\theta_{{\rm beam}}\Big)^{-2}, \quad \quad \bar w_X=\sum_c w_{X,c}\;,\\
&&\bar B_{l}=\f{1}{\bar w_X}\sum_c w_{X,c}B_{c,l}\,, \quad X\in(T,P)\\
&&B_{c,l}\simeq {\rm exp}\le\{\f{-l(l+1)\theta^2_{{\rm beam}}}{8\ln 2}\ri\}\;.\eea
In the above expressions, $B_l^2$ is the Gaussian beam window function
,$\theta_{{\rm beam}}$ is the beam's full-width at half-maximum
(FWHM), and $\sigma_X$ are the sensitivities.
In this paper we choose beams and sensitivities representative of
WMAP and \emph{Planck} experiments, following specifications from
\cite{Eisenstein:1998hr,Planck:2006aa}. For convenience we list
channels, beams
and noise levels we used in Tab.\ref{tab:experiment}.
\begin{table}
\caption{\label{tab:experiment}WMAP and Planck experiment}
\begin{ruledtabular}
\begin{tabular}{cccccccc}
Experiment & Frequency & $\theta_{{\rm beam}}$ & $\sigma_T$ & $\sigma_P$ \\
\hline
WMAP:		&	94	&	12.6 & 49.9 & 70.7 \\
			&   60  &   21.0 & 30.0 & 42.6 \\
			&   40  &   28.2 & 17.2 & 24.4 \\
\hline
Planck:		&   217 &   5.0  & 13.1 & 26.8 \\
			&   143 &   7.1  & 6.0  & 11.5 \\Junk:2012qt
			&   100 &   10.0 & 6.8  & 10.9 \\
			&   70  &   14.0 & 12.8 & 18.3 \\			
\end{tabular}
\end{ruledtabular}
\footnotetext{Frequencies in GHz. Beam size $\theta_{{\rm beam}}$ is the FWHM in
arcminutes. Sensitivities $\sigma_T$ and $\sigma_P$ are in $\mu K$ per FWHM beam.}
\end{table}

As shown in the main text, the likelihood of $B_0$ is highly
non-Gaussian, which makes forecasting $B_0$ error bars through a
Fisher matrix analysis unfeasible. On the other hand we still use the
Fisher matrix as a way to get a reasonable order of magnitude estimate
of the correlation between different parameters.

We jointly analyze $7$ cosmological parameters within the Hu-Sawicki
model ($A_s,n_s,\tau,\Omega_bh^2,\Omega_mh^2,h,B_0$). The first $6$
are the standard $\Lambda CDM$ parameters, while $B_0$ is the new
modified gravity parameter. Actually, as stated in Sec.\ref{IIB} the Hu-Sawicki model is charaterized by $2$ parameters ($B_0,n$), however, in our calculations we fix $n$ for a few specific values. The are various reasons for this choice: firstly, there exists a parameter degeneracy between $n$ and $B_0$; secondly, 
the current constraints on the parameter $n$ is rather weak \cite{Martinelli:2011wi} and the results are strongly dependent on the value of $n$. Given the above considerations, we fix $n$ for a few specific values in the following numerical analysis, i.e. we can deal with the Hu-Sawicki model with different $n$'s as different models.

Our Fisher matrix results are summarized in Table~\ref{tab:joint_analysis}. 
The first six rows are obtained for the Hu-Sawicki model with $n=4$. 
Actually, we find that the dependence of these standard parameters on $n$ is rather weak, i.e. 
we can obtain similar results for other values of $n$.
As a numerical check, we verified that our results for the $6$
``standard'' parameters match well both WMAP-7yr \cite{WMAP7yr} and
Planck blue book results~\cite{Planck:2006aa}. We find no significant
degeneracies between $B_0$ and other parameter, thus justifying the
simple 1-parameter likelihood approach we use in the main text to
forecast $B_0$ error bars

\begin{table}
\caption{\label{tab:joint_analysis}Fisher matrix joint analysis results ($68\%$CL) on lensed power spectrum for the Hu-Sawicki model}
\begin{ruledtabular}
\begin{tabular}{cccccccc}
CP & Fiducial value\footnotemark[1] & WMAP & PLANCK \\
\hline
$A_s$
		&	$2.43\times10^{-9}$		&	$9.61\times10^{-11}$		&	$3.46\times10^{-11}$ \\	
$n_s$
		&	$0.963$				&	$1.15\times10^{-2}$		&	$3.38\times10^{-3}$  \\
$\tau$
		&	$0.088$				&	$1.58\times10^{-2}$		&	$3.92\times10^{-3}$  \\
$\Omega_bh^2$
		&	$0.0226$			&	$5.53\times10^{-4}$		&	$1.47\times10^{-4}$  \\
$\Omega_mh^2$
		&	$0.1109$			&	$5.31\times10^{-3}$		&	$1.23\times10^{-3}$  \\
$h$
		&	$0.70$				&	$1.72\times10^{-2}$		&	$6.20\times10^{-3}$  \\
$B_0 (n=4)$\footnotemark[2]
		&	$0$				    &	$1.81\times10^{-3}$				&	$8.33\times10^{-5}$  		 \\
$B_0 (n=1)$
		&	$0$				    &	$4.41\times10^{-4}$				&	$1.92\times10^{-5}$  		 \\		
\end{tabular}
\end{ruledtabular}
\footnotetext[1]{Fiducial values of the first six parameters are taken as WMAP7yr best-fit values.}
\footnotetext[2]{Results of the first six parameters include polarization contributions, while those of $B_0$
are only from $C_l^{TT}$ mode.}
\end{table}

Furthermore, in Tab. \ref{tab:B0Cross} we list the cross-correlation coefficients of several cosmological parameters with $B_0$ in Hu-Sawicki model with $n=4$. Our results show that all the cross-correlation coefficients are small. 

\begin{table}
\caption{\label{tab:B0Cross}Cross-correlation coefficients of the standard cosmological parameters with $B_0$ in Hu-Sawicki model ($n=4$)}
\begin{ruledtabular}
\begin{tabular}{cccccccc}
$B_0$-CP & cross-correlation coefficients\\
\hline
$A_s$
		&	$-7.9\times10^{-2}$		\\	
$n_s$
		&	$-0.12$			 \\
$\tau$
		&	$-8.0\times10^{-2}$			  \\
$\Omega_bh^2$
		&	$-0.18$		  \\
$\Omega_mh^2$
		&	$-4.7\times10^{-2}$		  \\
$h$
		&	$8.2\times10^{-2}$			  \\	
\end{tabular}
\end{ruledtabular}
\end{table}

\section{\label{App:example}Computation example of $\sigma^2_{\rm PB}$}
As an  example of the computation for the cross-likelihood, Eq. \ref{PBCross} we show 
explicitly the following term:
\bea\label{example}
&&\Big\la\hat C^{T}_l\hat C^{\phi T}_{l_2}\hat C^{{\rm T}}_{l_3}\Big\ra -\bar C^{T}_l\bar B^{{\rm ISW-L}}_{l_1l_2l_3}\nn\\
&=&\sum_{m',m'_2,m'_3,m,m_2,m_3}\f{\delta_{mm'}\delta_{m_2m'_2}\delta_{m_3m'_3}}{(2l+1)(2l_2+1)(2l_3+1)}\nn\\
&&\Big\la a^{T\ast}_{lm}a^{T}_{lm}\phi^{\ast}_{l_2m_2}a^{T}_{l_2m_2}a^{{\rm T}\ast}_{l_3m_3}a^{{\rm T}}_{l_3m_3}\Big\ra-\bar C^{T}_l\bar B^{{\rm ISW-L}}_{l_1l_2l_3}\nn\\
&=&\sum_{m',m'_2,m'_3,m,m_2,m_3}\f{\delta_{mm'}\delta_{m_2m'_2}\delta_{m_3m'_3}}{(2l+1)(2l_2+1)(2l_3+1)}\nn\\
&&\times\le\{4\contraction[1.5ex]{\Big\la}{a^{T\ast}_{lm}}{a^{T}_{lm}\phi^{\ast}_{l_2m_2}}{a^{T}_{l_2m_2}}
\bcontraction[2.5ex]{\Big\la a^{T\ast}_{lm}}{a^{T}_{lm}}{\phi^{\ast}_{l_2m_2}a^{T}_{l_2m_2}a^{{\rm T}\ast}_{l_3m_3}}{a^{{\rm T}}_{l_3m_3}}
\bcontraction[1.5ex]{\Big\la a^{T\ast}_{lm}a^{T}_{lm}}{\phi^{\ast}_{l_2m_2}}{a^{T}_{l_2m_2}}{a^{{\rm T}\ast}_{l_3m_3}}
\Big\la a^{T\ast}_{lm}a^{T}_{lm}\phi^{\ast}_{l_2m_2}a^{T}_{l_2m_2}a^{{\rm T}\ast}_{l_3m_3}a^{{\rm T}}_{l_3m_3}\Big\ra\ri.\nn\\
%
%
%
%
&&+\contraction{2\Big\la}{a^{T\ast}_{lm}}{a^{T}_{lm}}{\phi^{\ast}_{l_2m_2}}
\bcontraction[1.5ex]{2\Big\la a^{T\ast}_{lm}}{a^{T}_{lm}}{\phi^{\ast}_{l_2m_2}}{a^{T}_{l_2m_2}}
\contraction{2\Big\la a^{T\ast}_{lm}a^{T}_{lm}\phi^{\ast}_{l_2m_2}a^{T}_{l_2m_2}}{a^{{\rm T}\ast}_{l_3m_3}}{}{a^{{\rm T}}_{l_3m_3}}
2\Big\la a^{T\ast}_{lm}a^{T}_{lm}\phi^{\ast}_{l_2m_2}a^{T}_{l_2m_2}a^{{\rm T}\ast}_{l_3m_3}a^{{\rm T}}_{l_3m_3}\Big\ra\nn\\
&&+\contraction{4\Big\la}{a^{T\ast}_{lm}}{a^{T}_{lm}}{\phi^{\ast}_{l_2m_2}}
\bcontraction[1.5ex]{4\Big\la a^{T\ast}_{lm}}{a^{T}_{lm}}{\phi^{\ast}_{l_2m_2}a^{T}_{l_2m_2}}{a^{{\rm T}\ast}_{l_3m_3}}
\contraction{4\Big\la a^{T\ast}_{lm}a^{T}_{lm}\phi^{\ast}_{l_2m_2}}{a^{T}_{l_2m_2}}{a^{{\rm T}\ast}_{l_3m_3}}{a^{{\rm T}}_{l_3m_3}}
4\Big\la a^{T\ast}_{lm}a^{T}_{lm}\phi^{\ast}_{l_2m_2}a^{T}_{l_2m_2}a^{{\rm T}\ast}_{l_3m_3}a^{{\rm T}}_{l_3m_3}\Big\ra\nn\\
&&\le.+\bcontraction[3.5ex]{2\Big\la}{a^{T\ast}_{lm}}{a^{T}_{lm}\phi^{\ast}_{l_2m_2}a^{T}_{l_2m_2}a^{{\rm T}\ast}_{l_3m_3}}{a^{{\rm T}}_{l_3m_3}}
\bcontraction[2.5ex]{2\Big\la a^{T\ast}_{lm}}{a^{T}_{lm}}{\phi^{\ast}_{l_2m_2}a^{T}_{l_2m_2}}{a^{{\rm T}\ast}_{l_3m_3}}
\bcontraction[1.5ex]{2\Big\la a^{T\ast}_{lm}a^{T}_{lm}}{\phi^{\ast}_{l_2m_2}}{}{a^{T}_{l_2m_2}}
2\Big\la a^{T\ast}_{lm}a^{T}_{lm}\phi^{\ast}_{l_2m_2}a^{T}_{l_2m_2}a^{{\rm T}\ast}_{l_3m_3}a^{{\rm T}}_{l_3m_3}\Big\ra\ri\}\;,
\eea
where we have subtracted $3$ disconnected terms.

\end{document}